# Spontaneous non-ground state polariton condensation in pillar microcavities

**M. Maragkou[1], A. J. D. Grundy[1], E. Wertz[2], A. Lemaître[2], I. Sagnes[2], P. Senellart[2], J. Bloch[2], P. G. Lagoudakis[1]**

*1. School of Physics and Astronomy, University of Southampton, Southampton SO17 1BJ, United Kingdom*

*2. Laboratoire de Photonique et Nanostructures, LPN/CNRS, Route de Nozay, 91460 Marcoussis, France*

**We observe spontaneously driven non-ground state polariton condensation in GaAs pillar microcavities under non-resonant optical excitation. We identify a regime where the interplay of exciton-exciton and pair polariton scattering can lead to mode switching from non-ground state to ground state polariton condensation. A simple kinematic model satisfactorily describes the observed mode switching as each of the above scattering mechanisms becomes prevalent at different carrier densities.**

Polaritons in semiconductor microcavities are the admixture of an exciton and a cavity photon in the strong coupling regime.[1, 2] Due to their photon component, the de Broglie wavelength of polaritons is several orders of magnitude larger than that of atoms, allowing in principle for Bose Einstein condensation (BEC) even at room temperature.[3,4,5] However, unlike atoms, the polariton lifetime is limited by the photon cavity lifetime to a few picoseconds. Although the ultrashort lifetime prevents thermalisation with the host lattice, inter-particle interactions allow rapid relaxation and the formation of a macroscopically occupied ground state, usually referred to as a polariton condensate[6]. The non-equilibrium nature of polariton condensates, described by Imamoglu in 1996 in the context of an inversionless laser, distinguishes them from pure BEC[7] and it is only within the framework of non-equilibrium BEC that polariton condensates can be rigorously described.[8] Although in an infinite 2D system BEC formally cannot occur, polariton condensation is observed in nominally 2D microcavities due to the spatial localization of polaritons in the photonic disorder of the cavities[9,10,11]. The localization required for condensation can be controlled by engineering tunable potential traps[12,13], or by etching planar samples into microcavity pillars.[14,15,16]

Spontaneously occurring ground state condensates have been reported in atomic and solid state systems and are well described by the theory of equilibrium and non-equilibrium BEC. An interesting extension to this theory has predicted that BEC can occur in states other than the ground state, but to date there has been no experimental evidence of non-ground state BEC.[17] In this letter we demonstrate spontaneously occurring non-ground-state polariton condensation in GaAs/AlGaAs pillar microcavities under non-resonant optical excitation. We show that by tuning the level separation of the polariton energy states in pillar microcavities we can control the mode switching between non-ground-state and ground-state polariton condensation. A kinematic model is introduced where the interplay of exciton-exciton and pair polariton scattering can adequately describe the observed mode switching as each of the above scattering mechanisms becomes prevalent at different carrier densities.

In pillar microcavities the contrast of the air-semiconductor index of refraction laterally confines the photon mode, which combined with the vertical confinement imposed by a pair of Bragg mirrors, produces a ladder of discrete 0D photon states determined predominantly by the size and shape of the pillar.[18,19] In the strong coupling regime the dressed states of the system form a sequence of discrete energy states, while the broken translational invariance of photon modes removes the restriction of momentum conservation from polariton-polariton scattering -usually referred to as pair polariton scattering. The latter allows for efficient polariton relaxation, removing the relaxation bottleneck usually encountered in planar microcavity structures that impedes thermalisation and polariton condensation.[20] Under non-resonant optical excitation, relaxation from the exciton reservoir to the polariton states in the linear regime is predominantly driven by exciton-exciton scattering, while pair polariton scattering contributes to relaxation between the discrete polariton states. The interplay of exciton-exciton and pair polariton scattering defines which of the polariton states will first reach occupancy of one and thus drive the system in the non-linear regime where polariton condensation dominates the dynamics.

Our sample, previously described in Ref. [14] consists of a $\lambda/2$ $Ga_{0.05}Al_{0.95}As$ cavity surrounded by 2 $Ga_{0.05}Al_{0.95}As/Ga_{0.80}Al_{0.20}As$ Bragg mirrors with 26 and 30 pairs in the top and bottom mirrors respectively. In order to achieve a large Rabi splitting and therefore maintain strong coupling between the exciton and the cavity mode three sets of four 7 nm GaAs quantum wells have been inserted in the centre and the first antinodes of the electromagnetic field within the structure. The sample is excited non-resonantly using 150 fs transform limited pulsed excitation with a repetition rate of 80 MHz. A microscope objective, aligned normal to the sample surface is used to both focus the excitation beam to a spot of 3 μm and collect photoluminescence. The sample is cooled to 7 K in a continuous flow cold finger cryostat. In order to achieve efficient excitation the laser is tuned to 1.70 eV, corresponding to the second Bragg mode of the system above the stop band, where the reflectivity is at a minimum, as shown with an arrow in Fig. 1(a). A spectrometer coupled to a streak camera is used to simultaneously record the temporal and energy dynamics of the photoluminescence.

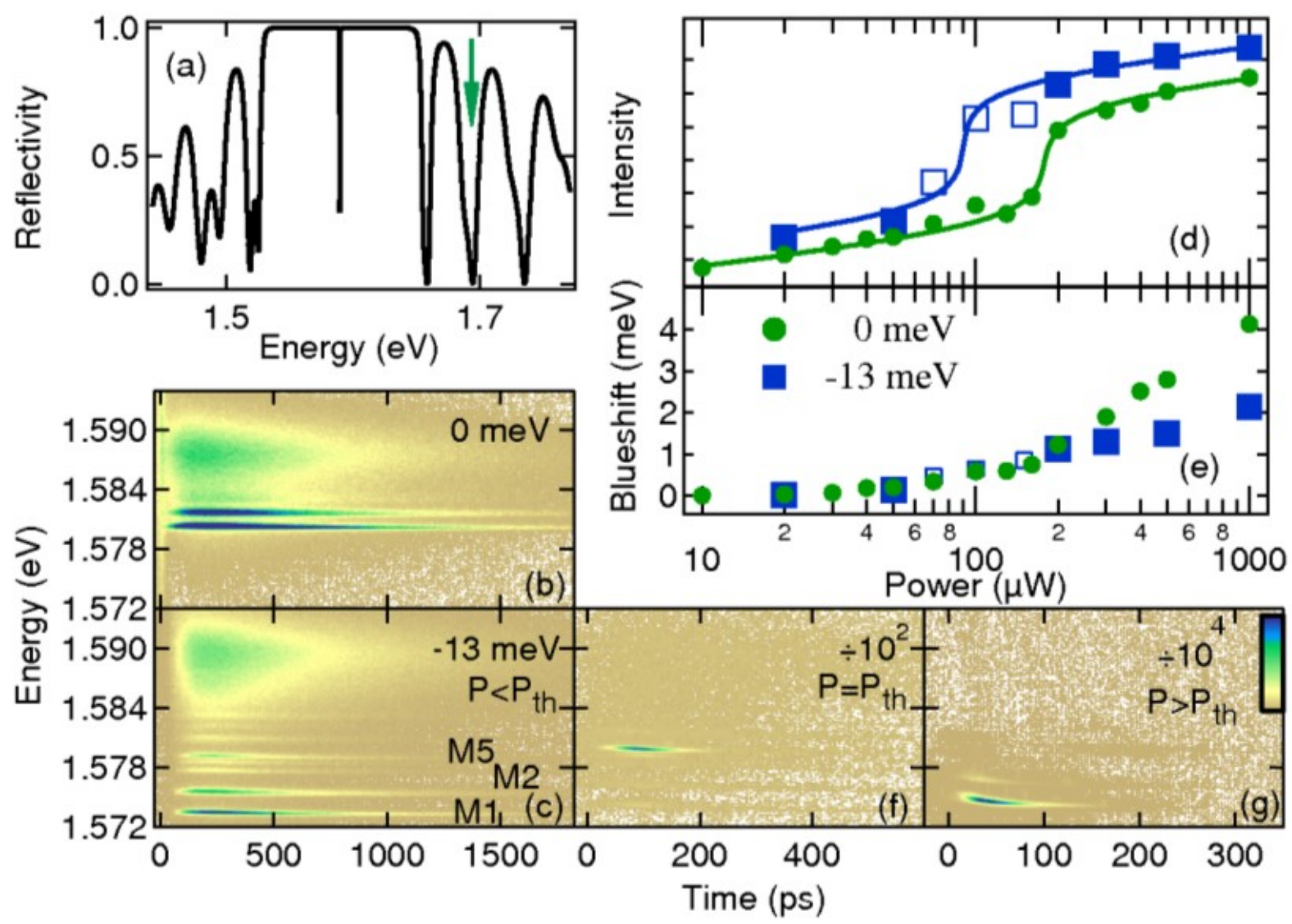

**FIG. 1**. (a) Reflectivity spectrum of the sample; the arrow indicates the energy of excitation. (b) Below threshold time and spectrally resolved photoluminescence from a 3.2 μm side square pillar at 0 meV and (c) -13 meV detuning conditions. (d) Power dependence of the emission intensity and (e) the energy blueshift that occurs for a 3.2 μm square pillar at 0 meV (green dots) and -13 meV (blue squares) detuning. The hollow blue squares correspond to the higher energy mode that exhibits polariton condensation for a certain range of excitation powers. (c, f, g) Time and spectrally resolved photoluminescence from a 3.2 μm pillar at -13 meV detuning below threshold, at threshold, and above threshold respectively (false color images with blue signifying higher intensity).

Spectrally and time resolved photoluminescence is recorded for 3.2 μm side square pillars at zero, and at negative (-13 meV) exciton-photon detuning conditions shown in Fig. 1(b,c). The excitation intensity is kept below threshold, for both detuning conditions. In both cases the highest energy state corresponds to uncoupled quantum well exciton photoluminescence collected from the side of the pillars. The photoluminescence decay dynamics of the lower energy states is not defined by the cavity life time (~15 ps here) but from the relaxation dynamics to each state. For both detuning conditions we observe that the states energetically closer to the exciton reservoir show a faster decay as expected from the scaling of the exciton-exciton scattering strength. The power dependence of the emission intensity at -13 meV and at zero detuning is shown in Fig. 1(d), with hollow markers indicating the regime where emission is strongest from the higher energy mode. We can confirm that emission occurs in the strong coupling regime by the continuous relative blue shift of the involved polariton states shown in Fig. 1(e). For comparison we have also plotted in the same figure the blue shift of the ground polariton state at zero detuning. At -13 meV the higher occupied energy state is switching from the ground polariton state to a higher energy state and back to the ground polariton state. The spontaneous mode switching occurs at the threshold to the non-linear regime rendering this observation interesting in the framework of spontaneously occurring non-ground state polariton condensation. Non-ground state polariton condensation was previously observed in pillar microcavities under non-uniform excitation conditions, where edge excitation of a pillar was used to trigger stimulated scattering towards polariton modes with matching field spatial distribution. Here, the pillars are uniformly illuminated and thus non-ground state polariton condensation is not driven by the excitation beam profile but is spontaneously occurring. We also note that mode switching could not be observed for any of the other available exciton-photon detuning conditions (-6 meV, 0 meV, +3.5 meV) under uniform excitation conditions.

Polaritons in pillar microcavities are not in thermal equilibrium, thus we attempt to understand the dynamics that lead to this unusual occurrence by studying the transient dynamics of polariton relaxation at the detuning condition where non-ground state polariton condensation is observed. Hereafter we use M1 to refer to the lowest energy polariton mode, M2 to the second lowest polariton mode and M5 to refer to mode where non-ground state polariton condensation occurs. The naming convention is shown in Fig. 1(c). We perform a power dependence of the non-resonant optical excitation and measure the photoluminescence decay dynamics of the different polariton modes as shown in Fig. 1(c, f, g). In the linear regime (Fig. 1(c)) the photoluminescence decay is of the order of several hundred picoseconds, dominated by the polariton relaxation mechanisms described above. At threshold we observe that polariton condensation occurs at the higher energy mode M5. The transition to the non-linear regime is observed both by the power dependence of the emission intensity (Fig. 1(d)), the collapse of the photoluminescence to a single mode and the reduction of the decay time by an order of magnitude (Fig. 1(f)). At excitation power three times above threshold we observe that polariton condensation switches to the lowest polariton mode, M1 (Fig. 1(g)). At this excitation power past polariton condensation we can resolve the repopulation of M5 at approximately 200 ps from the residual exciton reservoir (Fig. 1(g)). We also notice that although polariton condensation occurs at the lowest polariton mode M1, a blue shift occurs for all polariton states.

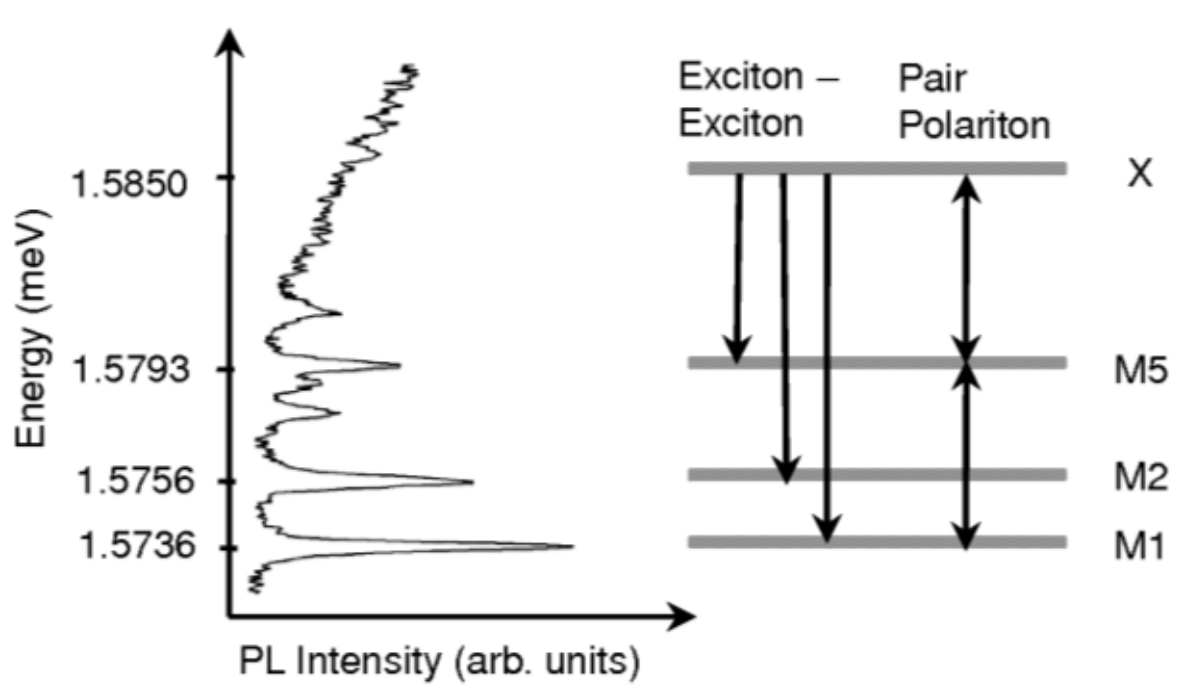

**FIG. 2**. Below threshold photoluminescence spectrum from a 3.2μm square pillar at δ=-13 meV detuning and schematic representation of the energy transitions between the exciton reservoir and polariton states M1, M2 and M5. Arrows indicate exciton-exciton and pair-polariton scattering between the states.

For polariton condensation to switch from a higher energy state (M5) to a lower energy state (M1) there needs to be a scattering mechanism that will populate M1 while depopulating M5. In Fig. 2 we show a photoluminescence spectrum in the linear regime and annotate the three polariton modes that dominate the observed dynamics and the relaxation channels coupling them. Other than exciton-exciton scattering, pair polariton scattering can occur for any three equidistant in energy modes, and at -13 meV detuning this process is indeed possible between M5, M1 and the low energy tale of the exciton reservoir. We expect that when M1 reaches occupation larger than one, final state stimulation of pair polariton scattering occurs, rapidly depleting M5 and therefore switching polariton condensation to M1. Here we note that pair polariton scattering in pillars does not require conservation of the in-plane wavevector and therefore the exciton states populated from pair polariton scattering will be distributed across a large range of wavevectors on the exciton dispersion. This can explain the absence of any observable enhancement of the emission from these exciton states within the light cone.

We can describe the observed dynamics quantitatively by considering a simple kinetic model that couples exciton reservoir and polariton modes through exciton-exciton, and pair polariton scattering. The following rate equations describe the population, $n_i$, for each of the states:[21,22,23]

$$\begin{aligned}\frac{\partial n_x}{\partial t} &= -\Gamma_X n_X - \sum_3 A {n_x}^2 (C_j n_j + 1) \\ &+ B(-C_{M1} n_{M1} D n_x (C_{M5} n_{M5} + 1)^2 \\ &+ C_{M5}^2 {n_{M5}}^2 (C_{M1} n_{M1} + 1)(D n_x + 1)) + f(t)\end{aligned} \qquad (1)$$

$$\begin{aligned}\frac{\partial n_j}{\partial t} &= -\Gamma_j n_j + A {n_x}^2 (C_j n_j + 1) \\ &+ B(-C_{M1} n_{M1} D n_x (C_{M5} n_{M5} + 1)^2 + \\ &C_{M5}^2 {n_{M5}}^2 (C_{M1} n_{M1} + 1)(D n_x + 1))\delta_{jM1} - \\ &2B(-C_{M1} n_{M1} D n_x (C_{M5} n_{M5} + 1)^2 + \\ &C_{M5}^2 {n_{M5}}^2 (C_{M1} n_{M1} + 1)(D n_x + 1))\delta_{jM5}\end{aligned} \qquad (2)$$

$$j = M1, M2, M5$$

where $n_x$ is the exciton population at the exciton reservoir, $A$, $B$ are the exciton-exciton and pair polariton scattering constants[24], $C_j$ is the exciton fraction of the $j$ state, $D$ is the fraction of the exciton reservoir that participates in pair-polariton scattering and $f(t)$ is the pumping term. The first term on the right hand side of the equations describes radiative decay from each state, where the decay rate $\Gamma j$ is extracted from the measured linewidth of each mode. The second term describes exciton-exciton scattering and the last terms describe the pair-polariton scattering. Energy conservation is required and limits pair polariton scattering to be between modes M1, M5 and the lower energy tale of the exciton reservoir (~4% of the total population), which are spaced equidistantly in energy.

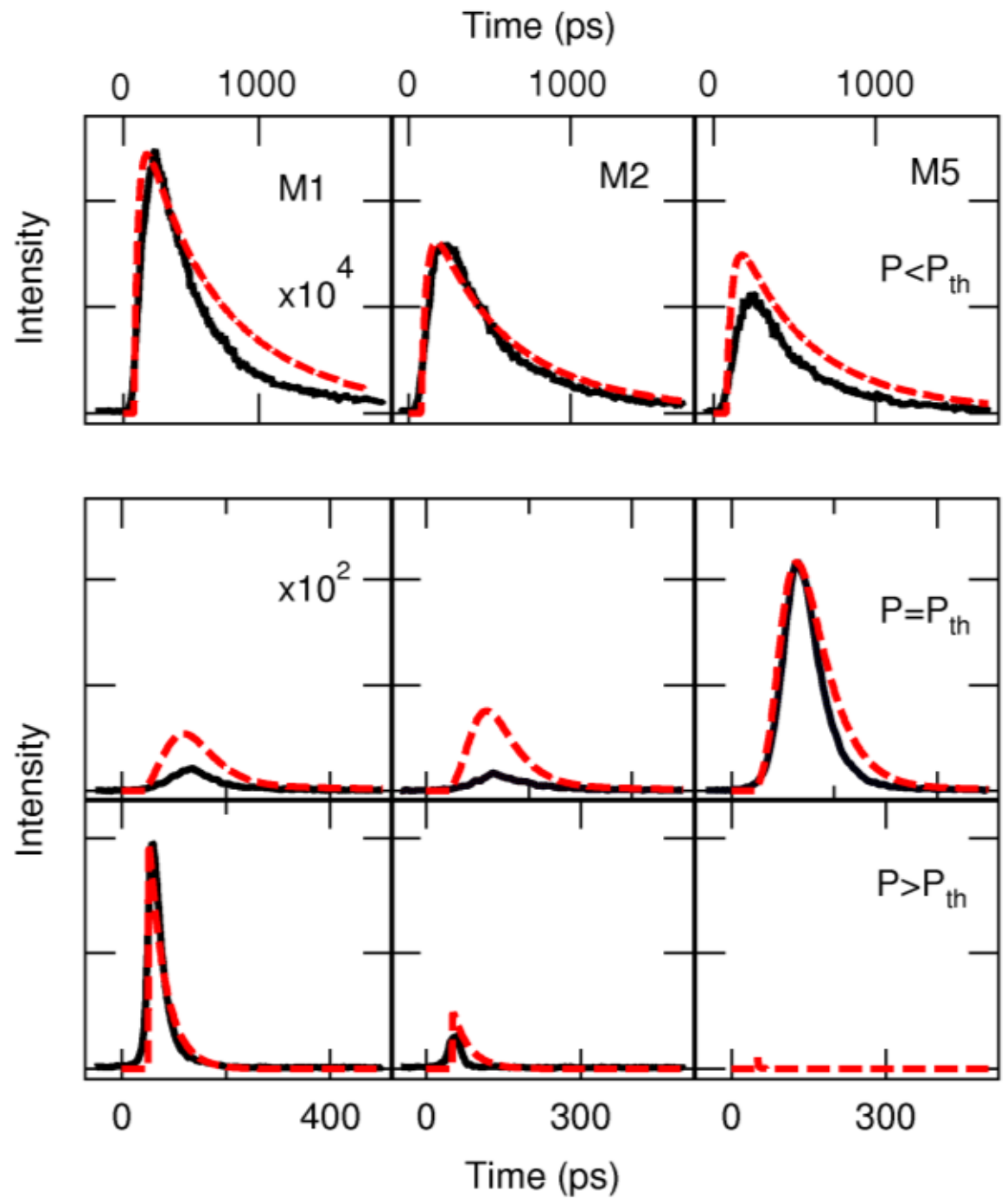

**FIG. 3** Comparison of the measured photoluminescence decay curves (solid black lines) and the fit of the kinematic model (dashed red lines). Columns correspond to different polariton states (M1, M2, M5 from left to right) and rows to different excitation power (below threshold, at threshold and above threshold from top to bottom).

The coupled differential equations are solved for different excitation powers and the transient populations of the three polariton modes scaled through the corresponding photon fraction are shown alongside the experimental data in Fig. 3. Each column corresponds to one polariton mode and each row to a different excitation power, matching those shown in Fig. 1(c, f, g). For low excitation powers (top panels) exciton-exciton scattering is the dominant relaxation mechanism and all modes exhibit PL of similar intensity, and decay times of hundreds of picoseconds, in agreement with the experimental data. At threshold (second row of panels in Fig. 3.), the higher energy state M5 is first to reach occupancy of one leading to a collapse of polaritons towards this state as we reach the threshold. As the pump power is increased above threshold pair polariton scattering dominates the dynamics and relaxation begins to favour the ground state again. The model replicates the experimental data and condensation switches to state M1 (third row in Fig. 3). This simple kinetic model adequately describes the most important features of the experiment, including condensation from different modes and changes in relaxation times of more than an order of magnitude.

In conclusion, the tunability of the relaxation mechanisms in pillar microcavities due to the discretisation of the energy states offers unique control over the state where polariton condensation occurs. We show that both ground and non-ground state polariton condensation is possible and the choice between the two is determined by the relaxation mechanism that is prevalent in the system for a particular carrier density. Within the framework of non-equilibrium BEC this is a manifestation of spontaneously occurring non-ground state BEC, the theory of which, although developed for trapped atoms, is currently work in progress for polaritons in semiconductor microcavities.

The authors would like to acknowledge Tomas Ostatnicky for helpful discussion on the modelling. This work was supported by EPSRC and EU FP7.

[24] The relative strength of the scattering constants A, B is 1:10.